\documentclass[10pt,a4paper,twocolumn,english,aps,prl,showpacs]{revtex4}
\usepackage[T1]{fontenc}
\usepackage[latin9]{inputenc}
\usepackage{graphicx}

\usepackage{babel}

\begin{document}

\title{Spin Superfluidity versus Solidity of Ultracold Polar Molecules}

\author{Hongyi Yu$^{1,2}$, W. M. Liu$^{1}$ }

\author{Chaohong Lee$^{2}$}

\altaffiliation{Corresponding author. Electronic addresses: chl124@physics.anu.edu.au; chleecn@gmail.com.}

\affiliation{$^{1}$Beijing National Laboratory for Condensed Matter Physics,
Institute of Physics, Chinese Academy of Sciences, Beijing 100190,
China \\
 $^{2}$Nonlinear Physics Centre and ARC Centre of Excellence for
Quantum-Atom Optics, Research School of Physics and Engineering, Australian
National University, Canberra ACT 0200, Australia }

\date{\today}
\begin{abstract}
We present a technique for engineering quantum magnets via ultracold
polar molecules in optical lattices and explore exotic interplay between
its spin superfluidity and solidity. The molecular ground and first
excited rotational states are resonantly coupled by a linearly polarized
microwave field. The spin-up (spin-down) states are presented by molecular
rotational states of electric dipole moment along (against) the coupling
field. By controlling the angle between the lattice direction and
the coupling field, the inter-site interaction can be tuned from antiferromagnetic
to ferromagnetic. Furthermore, the dipole-dipole interaction induces
an exotic interplay between spin superfluidity and solidity, and spin
supersolid phases may appear in mediate regions. 
\end{abstract}


\maketitle
In recent years, ultracold polar molecules in optical lattices have
stimulated great interests in both fundamental research and technology
application (for a recent review, see \cite{njpreview} and references
therein). Now, ultracold polar molecules such as LiCs and KRb can
be prepared into their rovibrational ground state \cite{prep,roviGS1,roviGS2,LiCs,KRb}.
Compared to ultracold atoms, ultracold polar molecules have some additional
promising features such as their complex rotational and vibrational
structures and strongly long-range anisotropic dipole-dipole interactions.
The high controllability and tunability of internal rotational structure
of polar molecules via external DC and AC electric fields makes them
ideal tools to design quantum logic gates for information processing
\cite{qgate1,qgate2,qgate3}. The strong and tunable dipole-dipole
interaction between ultracold polar molecules is very different from
the short-range interactions such as s-wave scattering. So that systems
of ultracold polar molecules provide an excellent opportunity for
simulating and exploring quantum phases in strongly correlated systems
of long-range interactions \cite{simu1,simu2,simu3}.

The superfluid - Mott insulator phase transition, which is well described
by a Bose-Hubbard model, has been experimentally demonstrated by manipulating
ultracold atoms in optical lattices \cite{MIexp}. It is suggested
that a supersolid phase with coexistence of diagonal and off-diagonal
long-range orders would appear in systems of off-site interactions
\cite{SSsuggest1,SSsuggest2}. To realize the off-site interactions,
it has proposed to use dipole-dipole interactions between dipolar
bosons \cite{dipoleboson} or promote bosons into higher bands \cite{highband}.
It seems that supersolid phase could be simulated by spinless dipolar
atoms in optical lattices, but the weak off-site interaction between
atoms hinders further progress. As bosons can be represented by spins,
supersolid states can also appear in quantum magnets, several numerical
studies on spin-dimer models have given positive conclusions \cite{dimer1,dimer2,dimer3}.
Different from the quantum magnets of conventional condensed matters,
the quantum magnets of ultracold polar molecules \cite{demler,zoller}
can be easily tuned in experiments.

In this Letter, we propose an alternative scheme to engineer quantum
magnets of long-range dipole-dipole interactions with ultracold polar
molecules in deep optical lattices. Usually, rotational excitations
of the molecules are much smaller than their electronic and vibrational
excitations. In the presence of a resonant AC electric field for first
two rotational states, the electronic and vibrational excitations
are almost absent, and so that molecular rotational states of dipole
along (against) the electric field can be used to present the spin-up
(spin-down) states. This spin presentation scheme differs from the
previous ones via electronic states \cite{zoller} or first excited
rotational state in the absence of an external electric field \cite{demler}.
Furthermore, our scheme provides a direct realization of spinor qubits
\cite{qgate1} rather than an equivalent mapping or a quantum projection.
The long-range interaction between polar molecules naturally brings
the exchange interaction between effective spins. Various quantum
phases including superfluid and supersolid in quantum spin-$\frac{1}{2}$
chain of one polar molecule per site are explored. By utilizing existed
experimental techniques, the possibility of preparing and manipulating
such kind of quantum magnets are briefly discussed.

\begin{figure}
\includegraphics[width=1\columnwidth]{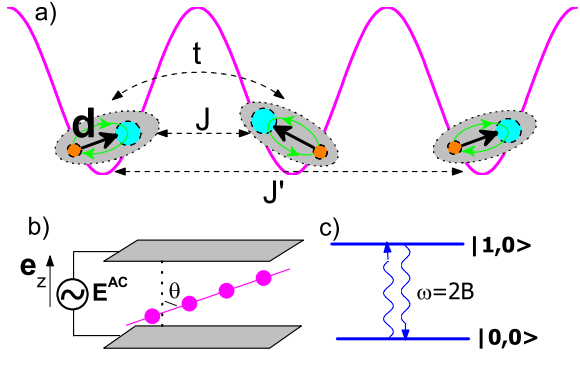} 

\caption{(Color online) (a) The schematic diagram for diatomic polar molecules
in an 1D optical lattice. (b) An AC electric field $\mathbf{E}^{\textrm{AC}}$
along $z$-axis is applied, where $\theta$ is the angle between the
$z$-axis and the lattice direction. (c) The AC field resonantly couples
two molecular rotational states $\left|0,0\right\rangle $ and $\left|1,0\right\rangle $,
i.e., $\omega=2B$.}

\label{molecule} 
\end{figure}

We consider an ensemble of ultracold polar molecules confined in a
deep optical lattice with one molecule per site, see Fig.~\ref{molecule}(a).
A single molecule with mass $m$ in such an optical potential obeys
Hamiltonian $H^{\textrm{mol}}=\hat{\mathbf{p}}^{2}/2m+H^{\textrm{opt}}+H^{\textrm{rot}}$.
Here, $\hat{\mathbf{p}}^{2}/2m$ and $H^{\textrm{opt}}$ are external
kinetic and potential energies for center-of-mass motions of the molecule,
respectively. $H^{\textrm{rot}}=B\hat{\mathbf{j}}^{2}$ is the internal
rotational motion characterized by the rotational constant $B$ for
electronic-vibrational ground state (for KRb, $B\approx1.1$ GHz \cite{KRb})
and the total angular momentum operator $\hat{\mathbf{j}}$. Introducing
quantum numbers $j$ and $M$ for the total angular momentum and its
$z$-component, the rotational eigenstates can be expressed as $|j,M\rangle$. 

In the presence of an AC electric field, $\mathbf{E}^{\textrm{AC}}(t)=E^{\textrm{AC}}e^{-i\omega t}\mathbf{e}_{z}+\textrm{c.c.}$,
the single-molecular Hamiltonian will have a part $H^{\textrm{AC}}(t)=-\hat{\mathbf{d}}\cdot\mathbf{e}_{z}E^{\textrm{AC}}e^{-i\omega t}+\textrm{h.c.}$
from the electric dipole interaction. Due to $\left\langle 0,0\right|\hat{d}_{z}\left|1,0\right\rangle =\left\langle 1,0\right|\hat{d}_{z}\left|0,0\right\rangle =d/\sqrt{3}$
(in which $\hat{d}_{z}=\mathbf{e}_{z}\cdot\hat{\mathbf{d}}$) \cite{simu3},
the AC electric field will mix $\left|0,0\right\rangle $ and $\left|1,0\right\rangle $.
The frequency $\omega$ is chosen as $2B$ which is the transition
frequency for $\left|0,0\right\rangle \to\left|1,0\right\rangle $,
see Fig.~\ref{molecule}(c). The AC electric field of such a frequency
can be generated by a microwave cavity. 

To keep $M$ unchanged, one has to use a microwave of linear polarization.
Due to transitions $\left|0,0\right\rangle \to\left|j,0\right\rangle $
of $j\geq2$ are far from resonant, the excitations to states $\left|j,0\right\rangle $
of $j\geq2$ can be ignored. Thus, the molecular states are coherent
superposition of $\left|0,0\right\rangle $ and $\left|1,0\right\rangle $.
Under a rotating wave approximation, the Hamiltonian in basis $\{\left|0,0\right\rangle ,\left|1,0\right\rangle \}$
reads \cite{simu2} \begin{eqnarray}
H^{\textrm{rot}}+H^{\textrm{AC}}=-\frac{1}{2}\left[\begin{array}{cc}
0 & \Omega\\
\Omega & 0\end{array}\right],\end{eqnarray}
with $\Omega=E^{\textrm{AC}}\left\langle 1,0\right|\hat{d}_{z}\left|0,0\right\rangle =E^{\textrm{AC}}d/\sqrt{3}$.
Introducing a new basis of `dipole-up' $\left|\uparrow\right\rangle =\frac{1}{\sqrt{2}}\left(\left|0,0\right\rangle +\left|1,0\right\rangle \right)$
and `dipole-down' $\left|\downarrow\right\rangle =\frac{1}{\sqrt{2}}(\left|0,0\right\rangle -\left|1,0\right\rangle )$,
one can obtain $H^{\textrm{rot}}+H^{\textrm{AC}}=-\frac{1}{2}\Omega\sigma^{z}$
with $\sigma_{z}$ denoting the $z$-component Pauli matrix. Obviously,
$\left\langle \uparrow\right|\hat{\mathbf{d}}\left|\uparrow\right\rangle =d\mathbf{e}_{z}/\sqrt{3}$,
$\left\langle \downarrow\right|\hat{\mathbf{d}}\left|\downarrow\right\rangle =-d\mathbf{e}_{z}/\sqrt{3}$
and $\left\langle \uparrow\right|\hat{\mathbf{d}}\left|\downarrow\right\rangle =\left\langle \downarrow\right|\hat{\mathbf{d}}\left|\uparrow\right\rangle =0$.
Thus the dipole-dipole interaction $H_{dd}^{ij}$ between molecules
$i$ and $j$ reads \begin{eqnarray}
H_{dd}^{ij}\! & = & \!\frac{\hat{\mathbf{d}}_{i}\cdot\hat{\mathbf{d}}_{j}-3(\hat{\mathbf{d}}_{i}\cdot\mathbf{e}_{ij})(\hat{\mathbf{d}}_{j}\cdot\mathbf{e}_{ij})}{|\mathbf{r}_{ij}|^{3}}\nonumber \\
\! & = & \! d^{2}\frac{1-3\cos^{2}\theta}{3|\mathbf{r}_{ij}|^{3}}\sigma_{i}^{z}\sigma_{j}^{z},\end{eqnarray}
where $\sigma_{i}^{z}$ is the $z$-component Pauli-matrix of molecule
$i$, $\mathbf{r}_{ij}=r_{ij}\mathbf{e}_{ij}$ is the position vector
between two molecules, and $\theta$ is the angle between the coupling
field direction $\mathbf{e}_{z}$ and the lattice direction $\mathbf{e}_{ij}$,
see Fig. \ref{molecule}(b).

Under a Born-Oppenheimer approximation, the system of ultracold polar
molecules in deep optical lattices can be described by \begin{eqnarray}
H\!\! & = & \!\!\frac{1}{2}\sum_{i,\sigma}U_{\sigma\sigma}n_{i\sigma}\left(n_{i\sigma}-1\right)+U_{\uparrow\downarrow}\sum_{i}n_{i\uparrow}n_{i\downarrow}\nonumber \\
\! & + & \!\!\sum_{i\neq j}H_{dd}^{ij}-t\sum_{\langle ij\rangle,\sigma}(c_{i\sigma}^{\dag}c_{j\sigma}+h.c.)-h\sum_{i}S_{i}^{z},\label{hal}\end{eqnarray}
with spin indices $\sigma$$(=\uparrow\mbox{or}\downarrow)$, lattice
indices $i$ and $j$, molecular creation (destruction) operators
$c_{i\sigma}^{\dag}$ ($c_{i\sigma}$), on-site interaction $U_{\sigma\sigma'}$,
nearest-neighbor tunneling $t$, molecular number operators $n_{i\sigma}=c_{i\sigma}^{\dag}c_{i\sigma}$,
and $z$-component spin operator $S_{i}^{z}=(c_{i\uparrow}^{\dag}c_{i\uparrow}-c_{i\downarrow}^{\dag}c_{i\downarrow})/2$.
Here, $H_{dd}^{ij}$ are off-site dipole-dipole interaction between
molecules, and the effective magnetic field strength $h$ is just
the Rabi frequency $\Omega$.

In the hard-core limit ($t\ll U_{\sigma\sigma'}$) of one molecular
per site ($n=n_{\downarrow}+n_{\uparrow}=1$), the `dipole-dipole'
interaction is $H_{dd}^{ij}=J_{ij}S_{i}^{z}S_{j}^{z}$ with $J_{ij}=4d^{2}(1-3\cos^{2}\theta)/3|(j-i)a|^{3}$,
where $a$ is the distance between two neighboring sites. By adjusting
$\theta$, $J_{ij}$ can be tuned from positive to negative continuously,
that is, the spin-spin interaction can be changed from antiferromagnetic
(AF) to ferromagnetic (FM). Due to the dipole-dipole interaction fast
decays according to $1/r^{3}$, we only take into account the nearest-neighbor
and the next-nearest-neighbor interactions with $J=4d^{2}(1-3\cos^{2}\theta)/3a^{3}$
and $J'=J/8$, respectively. Thus the Hamiltonian (\ref{hal}) is
equivalent to an anisotropic XXZ Heisenberg model \cite{xxz} \begin{eqnarray}
H\! & = & \!\pm J_{xy}\sum_{\langle ij\rangle}(S_{i}^{x}S_{j}^{x}+S_{i}^{y}S_{j}^{y})+J_{z}\sum_{\langle ij\rangle}S_{i}^{z}S_{j}^{z}\nonumber \\
 &  & \!+J'\sum_{\langle\!\langle ik\rangle\!\rangle}S_{i}^{z}S_{k}^{z}-h\sum_{i}S_{i}^{z},\label{hei}\end{eqnarray}
with $J_{z}=J+4t^{2}/U_{\uparrow\downarrow}-4t^{2}/U_{\downarrow\downarrow}-4t^{2}/U_{\uparrow\uparrow}$
and $J_{xy}=4t^{2}/U_{\uparrow\downarrow}$, the `$+$' and `$-$'
signs before $J_{xy}$ correspond to the cases of fermionic and bosonic
molecules, respectively. Here the $4t^{2}/U_{\sigma\sigma'}$ in $J_{z}$
and $J_{xy}$ are induced by spin-dependent collisions in second-order
tunneling processes. Below we shall only consider the bosonic case,
that is, $J=J_{z}-J_{xy}$. The fermionic case can be mapped onto
the bosonic case with a rotation of $\pi$ along $z$ axis on one
sublattice. The above XXZ Heisenberg chain (\ref{hei}) is equivalent
to a hard-core boson model \cite{inter,hc} \begin{eqnarray}
H & = & -\frac{J_{xy}}{2}\sum_{\left\langle ij\right\rangle }(a_{i}^{\dag}a_{j}+a_{j}^{\dag}a_{i})+J_{z}\sum_{\left\langle ij\right\rangle }n_{i}n_{j}\nonumber \\
 &  & +J'\sum_{\left\langle \!\langle ik\right\rangle \!\rangle}n_{i}n_{k}-\mu\sum_{i}n_{i},\label{hardcoreboson}\end{eqnarray}
 with $a_{i}=S_{i}^{-}$, $a_{i}^{\dag}=S_{i}^{+}$, $n_{i}=a_{i}^{\dag}a_{i}=S_{i}^{z}+\frac{1}{2}$
and $\mu=h+J_{z}+J'$.

\begin{figure}
\includegraphics[bb=17 25 300 160]{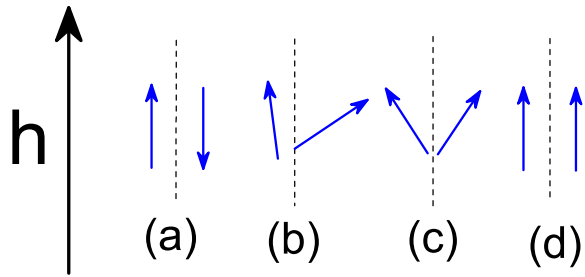} 

\caption{Various magnetic states of an 1D XXZ spin Hamiltonian (\ref{hei}).
(a) Antiferromagnetic state. (b) Intermediate state. (c) Spin-flop
state. (d) Paramagnetic state. }

\label{magstate} 
\end{figure}

The zero-temperature ground state of XXZ Heisenberg Hamiltonian can
be antiferromagnetic (AF), spin-flop (SF) or paramagnetic (PM) state
\cite{magnetic}, which correspond to mass density wave, superfluid
and Mott insulator states in the hard-core boson system, respectively.
It was also suggested that a first-order phase transition between
the antiferromagnetic and spin-flop states is splitted into two second-order
phase transitions by an `intermediate' (IM) state, in which diagonal
($S^{z}$) and off-diagonal ($S^{xy}$) long-range orders coexist,
which correspond to the coexistence of solid and superfluid order
in the hard-core boson system, thus it was also termed as supersolid.
These magnetic states can be characterized with various sublattice
magnetization as in Ref. \cite{inter}, see Fig.~\ref{magstate}.

The existence of diagonal long-range order is characterized by the
linear increase of static structure factor $S(0)$ or $S(\pi)$ with
the lattice size $N$ \cite{inter}, where $S(k)=\langle|\sum_{l}\exp(ikl)S_{l}^{z}|^{2}\rangle/N$
is the Fourier transformation of density-density or spin-spin correlations.
While the existence of off-diagonal long-range order is characterized
by the non-zero superfluid density or the spin-stiffness $\rho_{s}$,
which is defined as the second derivative of the ground-state energy
per site $E_{0}$ with respect to a twist $\Phi$ in the boundary
condition, see Ref. \cite{spinstiffness,sandvik_twist}.

\begin{figure}
\includegraphics[width=0.9\columnwidth]{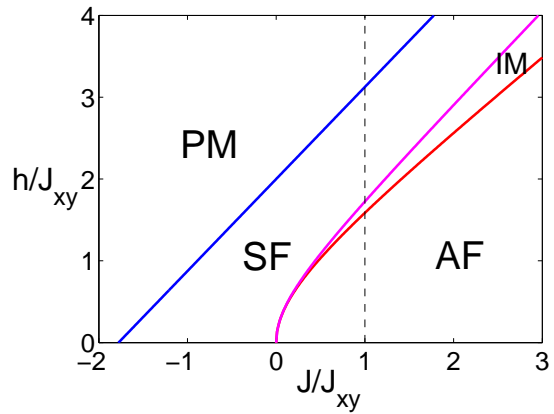} 

\caption{The $h$-$J$ phase diagram. The order parameters for a scan along
the dashed line of $J/J_{xy}=1$ are shown in Fig.~\ref{mag}.}

\label{phase} 
\end{figure}

To give a qualitative picture, we use a Gutzwiller variational ansatz
as the ground state wave function $|\Psi\rangle=\prod_{i}|\phi_{i}\rangle$.
The wave functions $|\phi_{i}\rangle$ for each site are expressed
in the basis of Fock states for hard-core bosns \cite{mf1,mf2}, i.e.,
$|\phi_{i}\rangle=f_{i}(0)|0\rangle_{i}+f_{i}(1)|1\rangle_{i}$. Using
the mean-field approximation, the system is divided into two intersecting
sublattices $A$ and $B$ with $f_{i}(0)=\sin\alpha$, $f_{i}(1)=\cos\alpha$
for $i\in A$, and $f_{j}(0)=\sin\beta$, $f_{j}(1)=\cos\beta$ for
$j\in B$. By minimizing the energy per site $E_{0}=\left\langle \Psi|H|\Psi\right\rangle /N$
with respect to $\alpha$ and $\beta$, the ground states in $J$-$h$
plane are shown in Fig.~\ref{phase}. In region of $J>0$, the phase
transition AF-IM-SF-PM occurs sequentially when $h$ gradually increases.
Because of the dipole-dipole interaction, values of $J'$ are nonzero
and supersolid states may appear in a certain region which increases
with $J$.

\begin{figure}
\includegraphics[width=0.9\columnwidth]{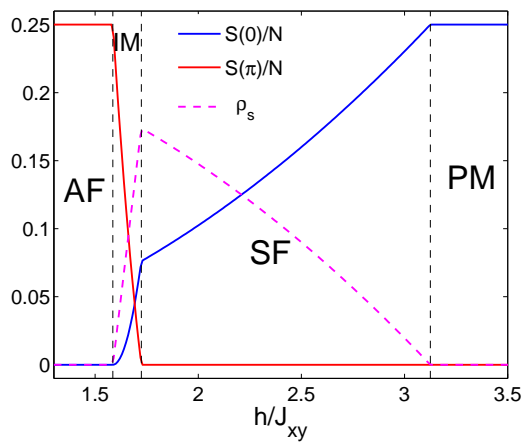} 

\caption{The mean-field results of static structure factors $S(0)$, $S(\pi)$
and superfluid density $\rho_{s}$ versus the effective magnetic field
$h$ at $J/J_{xy}=1$ and $J'=J/8$.}

\label{mag} 
\end{figure}

For a given mean-field ground state, one can obtain the static structure
factors $S(0)/N=(\cos2\alpha+\cos2\beta)^{2}/16$, $S(\pi)/N=(\cos2\alpha-\cos2\beta)^{2}/16$,
and the superfluid density $\rho_{s}=\frac{1}{4}\sin2\alpha\sin2\beta$
by using similar approaches in \cite{spinstiffness}. In Fig. \ref{mag}
we show how $S(0)$, $S(\pi)$ and $\rho_{s}$ depend on $h$ for
$J/J_{xy}=1$ and $J'=J/8$. From the figure we can see that in the
spin supersolid or intermediate state (region marked `IM' in the figure),
the diagonal long-range order $S(\pi)$ and off-diagonal long-range
order $\rho_{s}$ are all non-zero.

\begin{figure}
\includegraphics[width=1\columnwidth]{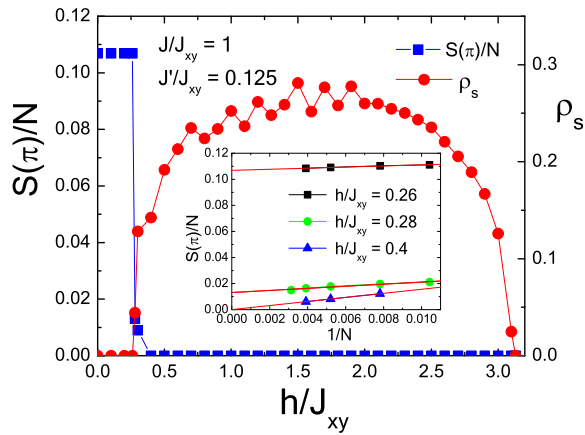} 

\caption{The QMC results of $S(\pi)$ and $\rho_{s}$ correspond to the mean-field
ones shown in Fig. \ref{mag}. The inset shows the finite size scaling:
$S(\pi)/N$ vs $1/N$ for different values of $h$.}

\label{MC} 
\end{figure}

We have also carried out stochastic series expansion (SSE) quantum
Monte Carlo (QMC) simulation \cite{sandvik_twist,sandvikSSE,directedloop,NNNimplementation}.
In our calculation, the inverse temperature is chosen as $4N$. The
superfluid density $\rho_{s}$ is measured via the fluctuations of
the winding number. For a system of $J/J_{xy}=1$ and $J'/J_{xy}=0.125$,
the dependence of $S(\pi)$ and $\rho_{s}$ on $h$ are shown in Fig.~\ref{MC}.
It shows that the QMC result is quantitatively different from the
mean field one. First, the region of AF state is greatly reduced because
of quantum fluctuation. Second, there is a sudden drop of $S(\pi)/N$
as $h$ increases, which shows a first order phase transition. We
find this sudden drop even appears when $J'/J_{xy}=0.25$ and $0.5$.
After the transition, the scaling behavior (see the inset of Fig.~\ref{MC})
shows $S(\pi)/N$ is still finite, it will gradually decrease to $0$
as $h$ increases further, thus a spin supersolid state appears.

Below we briefly discuss the experimental possibility. It has demonstrated
how to prepare polar molecules in optical lattices with one molecule
per site and cool them to their rovibrational ground states \cite{prep,roviGS1,roviGS2,LiCs,KRb}.
To ensure sufficiently large dipole-dipole interaction, one can choose
the heteronuclear diatom molecules such as LiCs and KRb which have
very large permanent electric dipole moments in an order of a few
Debye (for KRb in its singlet rovibrational ground state, $d\approx0.566$
Debye \cite{KRb}). The resonant coupling between states $\left|0,0\right\rangle $
and $\left|1,0\right\rangle $ can be achieved by a linearly polarized
microwave. The sign and strength of inter-site interaction can be
adjusted by controlling the wavelength of the optical lattice and
the angle between the lattice direction and the coupling field. The
effective magnetic field strength can be tuned by varying the coupling
strength, i.e., the Rabi frequency. 

In summary, by using ultracold polar molecules in optical lattices
and coupling their first two rotational states via an AC electric
field (microwave), we show an effective scheme for implementing long-range
dipole-dipole interaction. By controlling the optical lattice and
the coupling field, a quantum spin chain of long-range exchange interaction
can thus be simulated. Quantum phase transitions induced by tuning
the external field strength are explored. Experimental realization
of our scheme may give new insights into some open questions in condensed
matter physics including the competition between solid and superfluid
orders and the existence of supersolid. 

This work is supported by Australian Research Council (ARC), NSFC
(Grants No. 10874235, No. 10875039, No. 10934010 and No. 60978019)
and NKBRSFC (Grants No. 2006CB921400, No. 2009CB930704 and No. 2010CB922904).


\begin{thebibliography}{33}
\bibitem{njpreview} L. D. Carr \textit{et al.}, New J. Phys. ${\bf 11}$,
055049 (2009).

\bibitem{prep} T. Volz \textit{et al.}, Nature Phys. ${\bf 2}$,
692 (2006).

\bibitem{roviGS1} J. M. Sage \textit{et al.}, Phys. Rev. Lett. ${\bf 94}$,
203001 (2005).

\bibitem{roviGS2} J. G. Danzl \textit{et al.}, New J. Phys. ${\bf 11}$,
055036 (2009).

\bibitem{LiCs} J. Deiglmayr \textit{et al.}, Phys. Rev. Lett. ${\bf 101}$,
133004 (2008).

\bibitem{KRb} K.-K. Ni \textit{et al.}, Science ${\bf 322}$, 231
(2008).

\bibitem{qgate1} D. DeMille, Phys. Rev. Lett. ${\bf 88}$, 067901
(2002).

\bibitem{qgate2} C. Lee and E. A. Ostrovskaya, Phys. Rev. A ${\bf 72}$,
062321 (2005).

\bibitem{qgate3} S. F. Yelin \textit{et al.}, Phys. Rev. A ${\bf 74}$,
050301(R) (2006).

\bibitem{simu1} H. P. Buchler \textit{et al.}, Phys. Rev. Lett. ${\bf 98}$,
060404 (2007).

\bibitem{simu2} A. Micheli \textit{et al.}, Phys. Rev. A ${\bf 76}$,
043604 (2007).

\bibitem{simu3} G. Pupillo \textit{et al.}, arXiv:0805.1896.

\bibitem{MIexp} M. Greiner \textit{et al.}, Nature (London) ${\bf 415}$,
39 (2002).

\bibitem{SSsuggest1} C. Bruder \textit{et al.}, Phys. Rev. B ${\bf 47}$,
342 (1993).

\bibitem{SSsuggest2} G. G. Batrouni \textit{et al.}, Phys. Rev. Lett.
${\bf 74}$, 2527 (1995).

\bibitem{dipoleboson} K. Góral \textit{et al.}, Phys. Rev. Lett.
${\bf 88}$, 170406 (2002).

\bibitem{highband} V. W. Scarola \textit{et al.}, Phys. Rev. Lett.
${\bf 95}$, 033003 (2005).

\bibitem{dimer1} K.-K. Ng and T. K. Lee, Phys. Rev. Lett. ${\bf 97}$,
127204 (2006).

\bibitem{dimer2} P. Sengupta and C. D. Batista, Phys. Rev. Lett.
${\bf 98}$, 227201 (2007).

\bibitem{dimer3} N. Laflorencie and F. Mila, Phys. Rev. Lett. ${\bf 99}$,
027202 (2007).

\bibitem{zoller} A. Michel, G. K. Brennen, and P. Zoller, Nature
Phys. ${\bf 2}$, 341 (2006).

\bibitem{demler} R. Barnett, D. Petrov, M. Lukin, and E. Demler,
Phys. Rev. Lett. ${\bf 96}$, 190401 (2006).

\bibitem{xxz} L.-M. Duan E. Demler, and M. D. Lukin, Phys. Rev. Lett.
${\bf 91}$, 090402 (2003).

\bibitem{magnetic} D. P. Landau and K. Binder, Phys. Rev. B ${\bf 24}$,
1391 (1981).

\bibitem{inter} R. T. Scalettar \textit{et al.}, Phys. Rev. B ${\bf 51}$,
8467 (1995).

\bibitem{hc} C. Lee, Phys. Rev. Lett. ${\bf 93}$, 120406 (2004).

\bibitem{mf1} D. Jaksch \textit{et al.}, Phys. Rev. Lett. ${\bf 81}$,
3108 (1998).

\bibitem{mf2} D. S. Rokhsar and B. G. Kotliar, Phys. Rev. B ${\bf 44}$,
10328 (1991).

\bibitem{spinstiffness} A. Lüscher and A. M. Läuchli, Phys. Rev.
B ${\bf 79}$, 195102 (2009).

\bibitem{sandvik_twist} A. W. Sandvik, Phys. Rev. B ${\bf 56}$,
11678 (1997).

\bibitem{sandvikSSE} A. W. Sandvik, Phys. Rev. B ${\bf 59}$, R14157
(1999).

\bibitem{directedloop} O. F. Syljuåsen and A. W. Sandvik, Phys. Rev.
E ${\bf 66}$, 046701 (2002).

\bibitem{NNNimplementation} K. Louis and C. Gros, Phys. Rev. B ${\bf 70}$,
100410 (2004). 
\end{thebibliography}
\end{document}